\documentclass[reprint,aps,prb,floatfix,superscriptaddress,citeautoscript,a4paper,longbibliography]{revtex4-1}

\usepackage{amsmath, amsfonts, amssymb, amsbsy}
\usepackage{graphicx}
\usepackage{xspace}

\usepackage{hyperref}
\hypersetup{colorlinks=true, linkcolor=blue, citecolor=blue, urlcolor=blue, unicode=true}

\newcommand{\PbSnSe}{Pb\texorpdfstring{$_{0.77}$}{0.77}Sn\texorpdfstring{$_{0.23}$}{0.23}Se\xspace}
\newcommand{\PbSnSeTitle}{Pb\texorpdfstring{$_{\boldsymbol{0.77}}$}{0.77}Sn\texorpdfstring{$_{\boldsymbol{0.23}}$}{0.23}Se\xspace}
\newcommand{\PbSnXSe}{Pb\texorpdfstring{$_{1-x}$}{\ifpdfstringunicode{\unichar{"2081}\unichar{"208B}\unichar{"2093}}{1-x}}Sn\texorpdfstring{$_{x}$}{\ifpdfstringunicode{\unichar{"2093}}{x}}Se\xspace}
\newcommand{\PbSnXSeTitle}{Pb\texorpdfstring{$_{\boldsymbol{1-x}}$}{\ifpdfstringunicode{\unichar{"2081}\unichar{"208B}\unichar{"2093}}{1-x}}Sn\texorpdfstring{$_{\boldsymbol{x}}$}{\ifpdfstringunicode{\unichar{"2093}}{x}}Se\xspace}
\newcommand{\PbSnXTe}{Pb\texorpdfstring{$_{1-x}$}{\ifpdfstringunicode{\unichar{"2081}\unichar{"208B}\unichar{"2093}}{1-x}}Sn\texorpdfstring{$_{x}$}{\ifpdfstringunicode{\unichar{"2093}}{x}}Te\xspace}
\newcommand{\PbSnXTeTitle}{Pb\texorpdfstring{$_{\boldsymbol{1-x}}$}{\ifpdfstringunicode{\unichar{"2081}\unichar{"208B}\unichar{"2093}}{1-x}}Sn\texorpdfstring{$_{\boldsymbol{x}}$}{\ifpdfstringunicode{\unichar{"2093}}{x}}Te\xspace}

\begin{document}
\title{Topological crystalline insulator states in \PbSnXSeTitle}

\author{P.~Dziawa}
\affiliation{Institute of Physics, Polish Academy of Sciences, Aleja Lotnik\'{o}w 32/46, 02-668 Warsaw, Poland}
\author{B.~J. Kowalski}
\affiliation{Institute of Physics, Polish Academy of Sciences, Aleja Lotnik\'{o}w 32/46, 02-668 Warsaw, Poland}
\author{K.~Dybko}
\affiliation{Institute of Physics, Polish Academy of Sciences, Aleja Lotnik\'{o}w 32/46, 02-668 Warsaw, Poland}
\author{R.~Buczko}
\affiliation{Institute of Physics, Polish Academy of Sciences, Aleja Lotnik\'{o}w 32/46, 02-668 Warsaw, Poland}
\author{A.~Szczerbakow}
\affiliation{Institute of Physics, Polish Academy of Sciences, Aleja Lotnik\'{o}w 32/46, 02-668 Warsaw, Poland}
\author{M.~Szot}
\affiliation{Institute of Physics, Polish Academy of Sciences, Aleja Lotnik\'{o}w 32/46, 02-668 Warsaw, Poland}
\author{E.~\L{}usakowska}
\affiliation{Institute of Physics, Polish Academy of Sciences, Aleja Lotnik\'{o}w 32/46, 02-668 Warsaw, Poland}
\author{T.~Balasubramanian}
\affiliation{MAX IV Laboratory, Lund University, P.O. Box 118, 221 00 Lund, Sweden}
\author{B.~M. Wojek}
\affiliation{KTH Royal Institute of Technology, ICT Materials Physics, Electrum 229, 164 40 Kista, Sweden}
\author{M.~H. Berntsen}
\affiliation{KTH Royal Institute of Technology, ICT Materials Physics, Electrum 229, 164 40 Kista, Sweden}
\author{O.~Tjernberg}
\email{oscar@kth.se}
\affiliation{KTH Royal Institute of Technology, ICT Materials Physics, Electrum 229, 164 40 Kista, Sweden}
\author{T.~Story}
\email{story@ifpan.edu.pl}
\affiliation{Institute of Physics, Polish Academy of Sciences, Aleja Lotnik\'{o}w 32/46, 02-668 Warsaw, Poland}

\maketitle

{\bf Topological insulators are a class of quantum materials in which time-reversal symmetry, relativistic effects and an inverted band structure result in the occurrence of electronic metallic states on the surfaces of insulating bulk crystals. These helical states display a Dirac-like energy dispersion across the bulk bandgap, and they are topologically protected. Recent theoretical results have suggested the existence of topological crystalline insulators, a novel class of topological insulators in which crystalline symmetry replaces the role of time-reversal symmetry in ensuring topological protection~\cite{Fu-PhysRevLett-2011, Hsieh-NatCommun-2012}. In this study we show that the narrow-gap semiconductor \PbSnXSeTitle is a topological crystalline insulator for $\boldsymbol{x = 0.23}$. Temperature-dependent angle-resolved photoelectron spectroscopy demonstrates that the material undergoes a temperature-driven topological phase transition from a trivial insulator to a topological crystalline insulator. These experimental findings add a new class to the family of topological insulators, and we anticipate they will lead to a considerable body of additional research as well as detailed studies of topological phase transitions.}

The discovery of topological insulators (TI) is one of the most important recent developments in condensed-matter physics~\cite{Hasan-RevModPhys-2010, Qi-RevModPhys-2011, Hsieh-Nature-2008, Moore-Nature-2010, Xu-Science-2011, Koenig-Science-2007, Fu-PhysRevB-2007}. In these new quantum materials, time-reversal symmetry and strong relativistic (spin-orbit) effects require that the bulk insulating states are accompanied by metallic helical Dirac-like electronic states on the surface of the crystal. These surface states are encoded in topologically nontrivial wave functions of valence electrons and robustly resist non-magnetic disorder. The three-dimensional (3D) TIs Bi$_2$Se$_3$, Bi$_2$Te$_3$, and Bi$_{1-x}$Sb$_x$, along with their two-dimensional (2D) counterparts, which consist of HgTe/Hg$_{1-x}$Cd$_x$Te quantum wells, are considered to be model systems for this class of materials~\cite{Hasan-RevModPhys-2010, Qi-RevModPhys-2011, Hsieh-Nature-2008, Moore-Nature-2010, Xu-Science-2011, Koenig-Science-2007}. Novel quantum magnetotransport, magnetooptical and thermoelectric effects are expected in these materials, and the heterostructures composed of TIs and superconductors or ferromagnets~\cite{Hasan-RevModPhys-2010, Qi-RevModPhys-2011, Hsieh-Nature-2008, Moore-Nature-2010}.
 
In the search for new TI materials, the IV-VI semiconductors~\cite{NimtzAndSchlicht, Khokhlov, Pei-Nature-2011} PbTe, PbSe, and SnTe, as well as their substitutional solid solutions, \PbSnXTe and \PbSnXSe, have already been studied. However, these materials have been identified as trivial insulators in the topological classification of materials~\cite{Fu-PhysRevB-2007}. Several routes have been proposed to overcome this limitation by removing the four-fold valley degeneracy, e.g. by applying uniaxial strain~\cite{Fu-PhysRevB-2007} or exploiting the anisotropic energy quantisation of electrons confined at an interface~\cite{Buczko-PhysRevB-2012}. 

\begin{figure}
\centering
\includegraphics[width=.9\columnwidth]{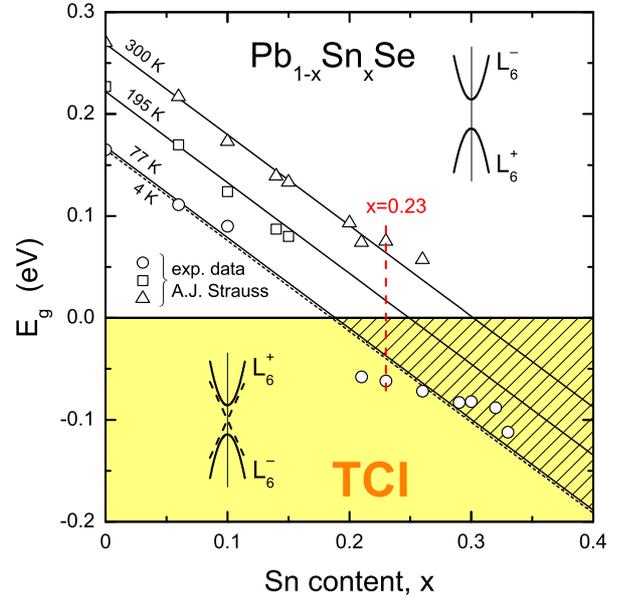}
\caption{{\bf \PbSnXSeTitle alloys as topological crystalline insulators.} The composition dependence of the bandgap of \PbSnXSe at various temperatures. The experimental data at $T=300$~K, $195$~K and $77$~K are from infrared absorption studies, and the $p$-$n$ junction laser emission studies of Strauss et al.~\cite{Strauss-PhysRev-1967}. The broken line for very low temperatures is an extrapolation based on both the known bandgap of PbSe ($E_\mathrm{g}=0.165$~eV at $T=4$~K) and the composition dependence parameter $\mathrm{d}E_\mathrm{g}/\mathrm{d}x=-0.89$~eV~\cite{Strauss-PhysRev-1967}. Positive bandgaps correspond to topologically trivial materials with PbTe-like or PbSe-like band symmetries. The materials with negative bandgaps (the yellow region) are topological crystalline insulators (TCI), with the bulk bandgap being open but having the (SnTe-like) inverted symmetry of the conduction and valence bands~\cite{Hsieh-NatCommun-2012}. The yellow-hatched region shows the composition and temperature ranges in which the TCI states exist in the \PbSnXSe alloy at ambient pressure.}
\label{fig:1}
\end{figure}

\begin{figure*}
\centering
\includegraphics[width=.38\textwidth]{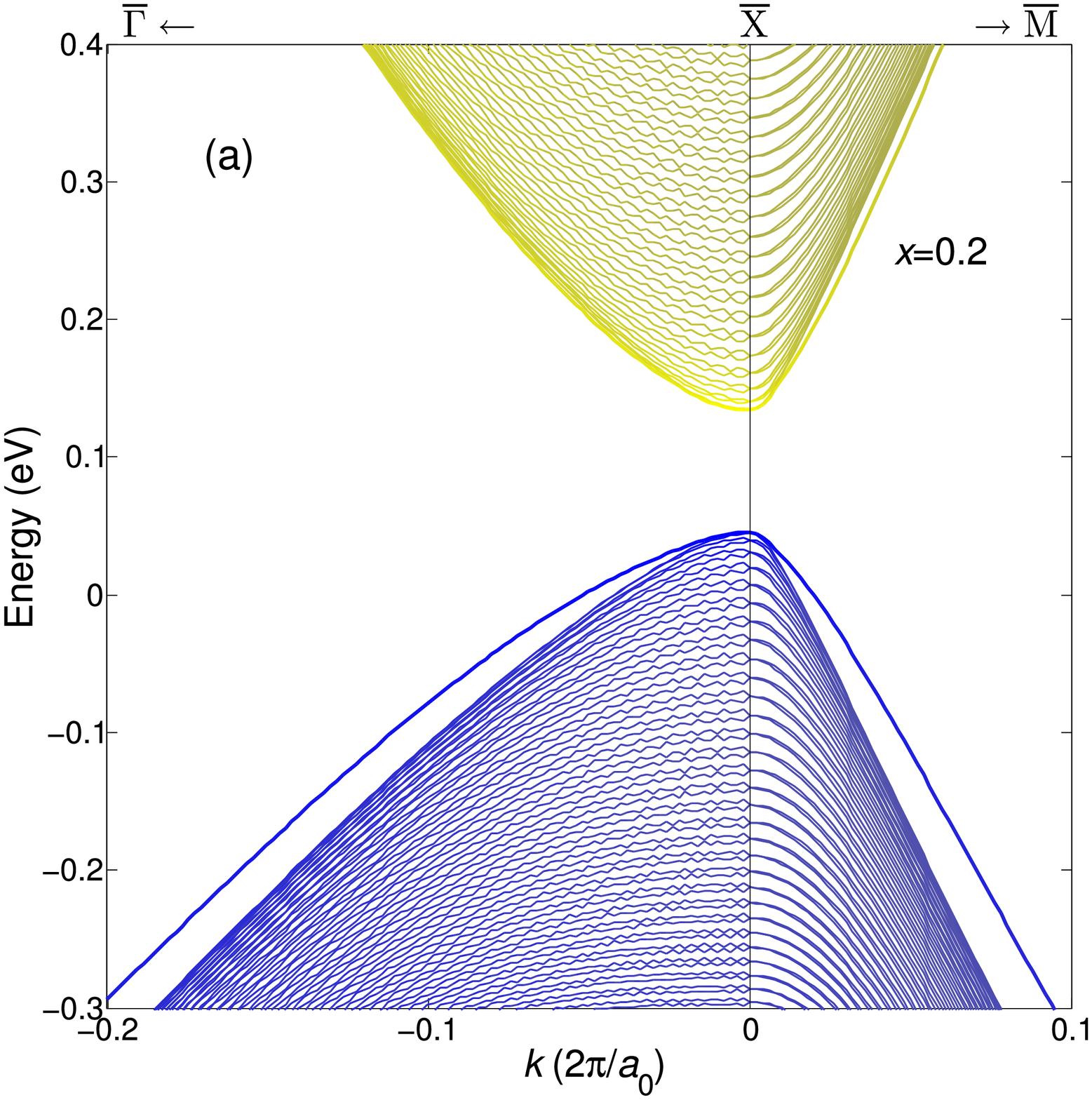}\hspace*{3mm}
\includegraphics[width=.38\textwidth]{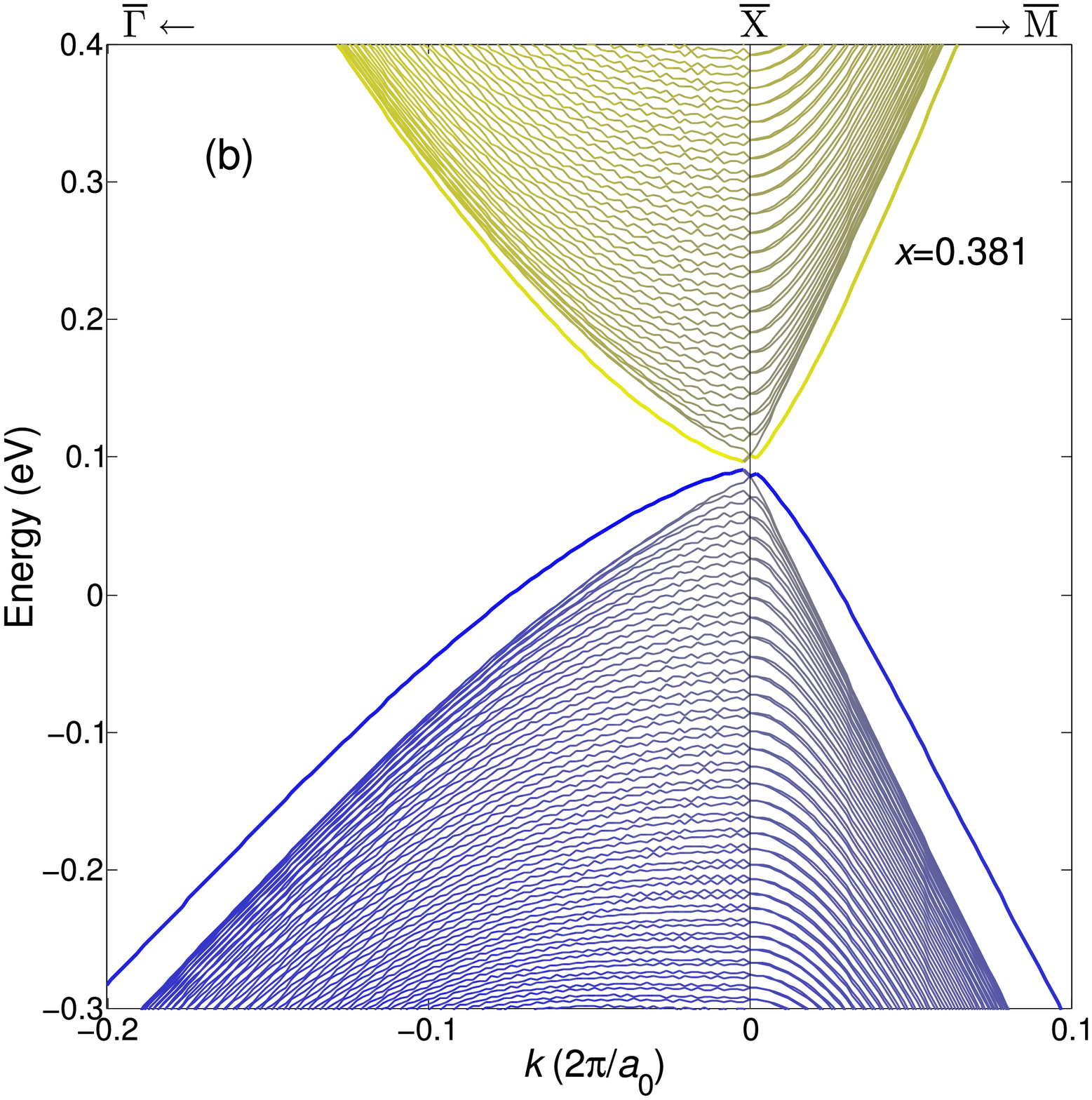}\\[3mm]
\includegraphics[width=.38\textwidth]{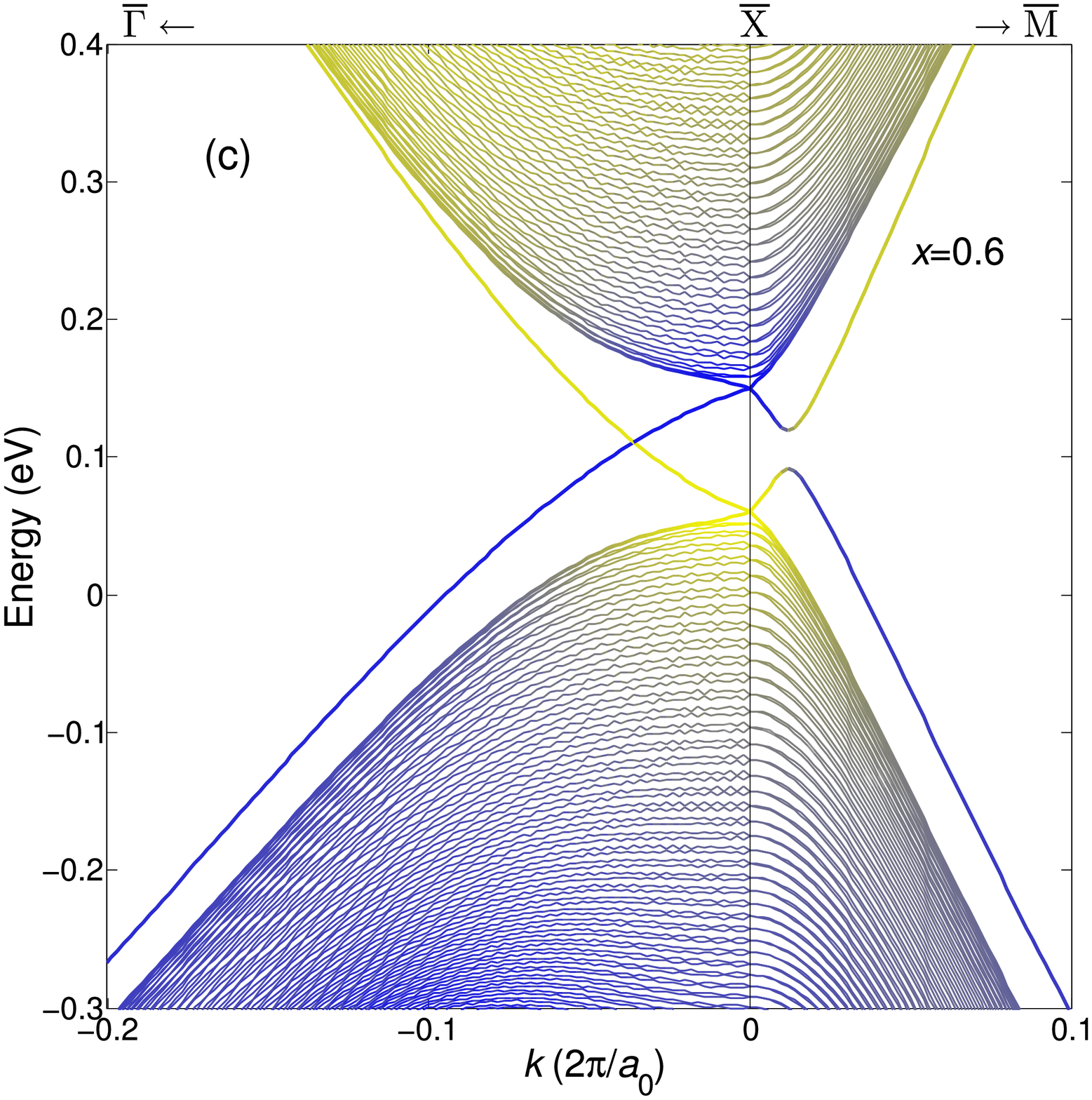}\hspace*{3mm}
\includegraphics[width=.38\textwidth]{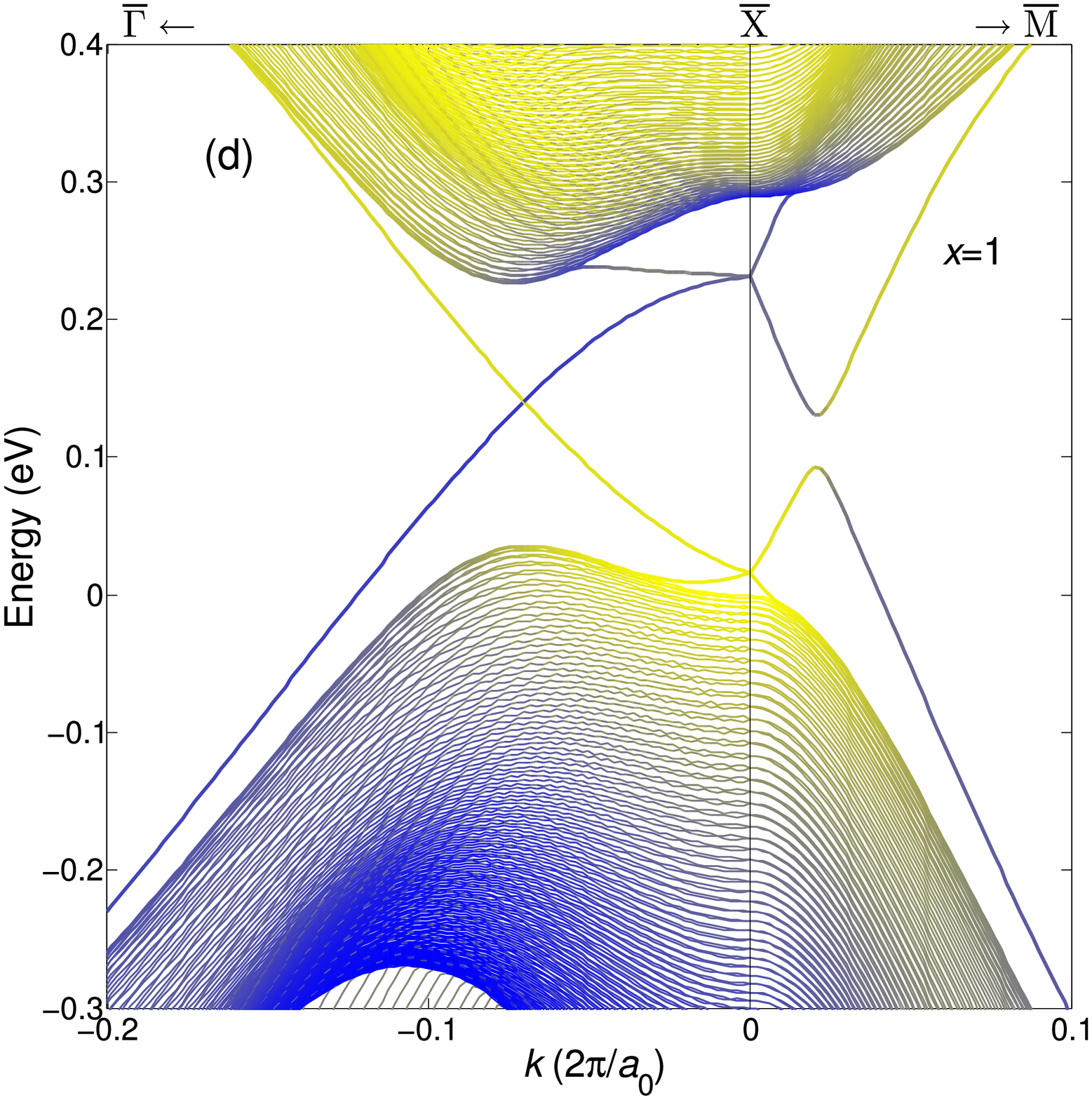}
\caption{{\bf Band structure calculations of \PbSnXTeTitle.} Theoretical tight-binding model calculations (in the virtual crystal approximation) of the band structure of a $280$~monolayer (i.e. about $90$~nm) thick slab of \PbSnXTe substitutional alloys with varying Sn content. \textbf{a}, Trivial insulator case ($x=0.2$). \textbf{b}, Zero bulk bandgap case ($x=0.381$). \textbf{c}, TCI case ($x=0.6$). \textbf{d}, SnTe TCI case ($x=1$). The densely packed lines represent the bulk states and the separated single lines correspond to the surface states. The wave functions of the surface states decay exponentially with the distance from the surface. The lines' colour changes from yellow to blue depending on the cation (yellow) and anion (blue) $p$-type orbitals dominant contribution to the state's wave function. Above the critical Sn composition $x_\mathrm{c}$ these contributions are exchanged between the top of the valence and the bottom of the conduction bands (band inversion).}
\label{fig:2}
\end{figure*}

Recent theoretical studies have proposed the idea of a new class of TIs, namely, topological crystalline insulators (TCI), in which specific crystalline symmetries warrant the topological protection of metallic surface states~\cite{Fu-PhysRevLett-2011, Hsieh-NatCommun-2012}. The first material that has been theoretically identified as a TCI is SnTe~\cite{Hsieh-NatCommun-2012}. Unfortunately, SnTe crystals are known to be heavily $p$-type, due to an exceptionally high concentration of electrically active Sn vacancies. The TCI states are not occupied~\cite{Littlewood-PhysRevLett-2012} and thus difficult to observe experimentally. Conversely, in the \PbSnXTe and \PbSnXSe alloys, the chemical potential can easily be tuned during the crystals' growth or annealing to yield $n$-type or $p$-type conductivity, which makes them more suitable for experimental investigations concerning the TCI state. The substitution of Sn for Pb in these alloys strongly changes the relativistic effects and results in a compositional evolution of their band structures, as presented in Figure~\ref{fig:1} for the selenide materials system~\cite{Strauss-PhysRev-1967, Dimmock-PhysRevLett-1966}. An increasing Sn content leads to a closing of the bulk bandgap at a specific critical alloy composition $x_\mathrm{c}$. For higher Sn contents, the bandgap opens again, with the parity of the electronic states at the band edges having been reversed, i.e., the situation is analogous to SnTe. In addition, Figure~\ref{fig:1} indicates that for a Sn content of $0.18 \leq x \leq 0.3$, a temperature-driven transition between a normal state and an inverted bandgap state occurs at a critical temperature $T_\mathrm{c}$. \PbSnXSe both crystallises in the rock-salt crystal structure and has a direct electronic bandgap ($E_\mathrm{g} \leq 0.29$~eV), which is located at four equivalent $L(111)$ points on the edge of the Brillouin zone~\cite{NimtzAndSchlicht, Khokhlov}. A strong indicator of \PbSnXSe being a TCI would be the formation of Dirac-like surface states that cross the bandgap. This phenomenon would only occur in the inverted bandgap state and therefore for $T<T_\mathrm{c}$.

We applied angle-resolved photoemission spectroscopy (ARPES) to determine experimentally the surface electronic band structure for a temperature range covering the band inversion. We found strong evidence for the TCI state in \PbSnSe at temperatures below the band inversion temperature $T_\mathrm{c}$. Additionally, a temperature-controlled topological phase transition from a trivial insulator ($T>T_\mathrm{c}$) to the TCI state ($T<T_\mathrm{c}$) was observed. We note that while the TCI state is topologically protected by crystalline symmetries~\cite{Fu-PhysRevLett-2011, Hsieh-NatCommun-2012}, the spin-orbit interaction is important in determining the actual band ordering and the details of the electronic structure in PbTe- and PbSe-based semiconductors and it is therefore expected to contribute to spin-polarisation effects and backscattering protection of the TCI surface states in these materials similar to what is observed in conventional TIs.

To verify the applicability of the TCI concept to the \PbSnXTe and \PbSnXSe substitutional alloys, we theoretically studied their electronic structures and their band inversions when $x=x_\mathrm{c}$. Tight-binding calculations were performed for $(001)$-oriented $280$-monolayer-thick crystal slabs of PbTe, SnTe, and \PbSnXTe alloys, with the Sn content varying across $x_\mathrm{c}$. The tight binding parameters for PbTe and SnTe were taken from Lent et al.~\cite{Lent-SuperlattMicrostruct-1986}. The analysis of the substitutional alloys ($0 < x < 1$) was performed using the virtual crystal approximation (VCA), which provides a good description of the band inversion~\cite{Mitchell-PhysRev-1966, Kriechbaum-PhysRevB-1984}. In contrast to SnTe, which has the same structure as rock-salt, SnSe crystallises in an orthorhombic structure, and the tight binding parameters for rock-salt SnSe crystals are not known. Still, in the Sn content range $x \leq 0.4$, the alloys \PbSnXTe and \PbSnXSe share the same rock-salt crystal structure, and their electronic bands have a notably similar symmetry. The band-structure parameters display the same dependence on temperature, pressure and chemical composition. Therefore, all of the qualitative conclusions of the theoretical analysis carried out for the \PbSnXTe alloy remain valid for the \PbSnXSe alloy.

\begin{figure*}
\centering
\includegraphics[width=.9\textwidth]{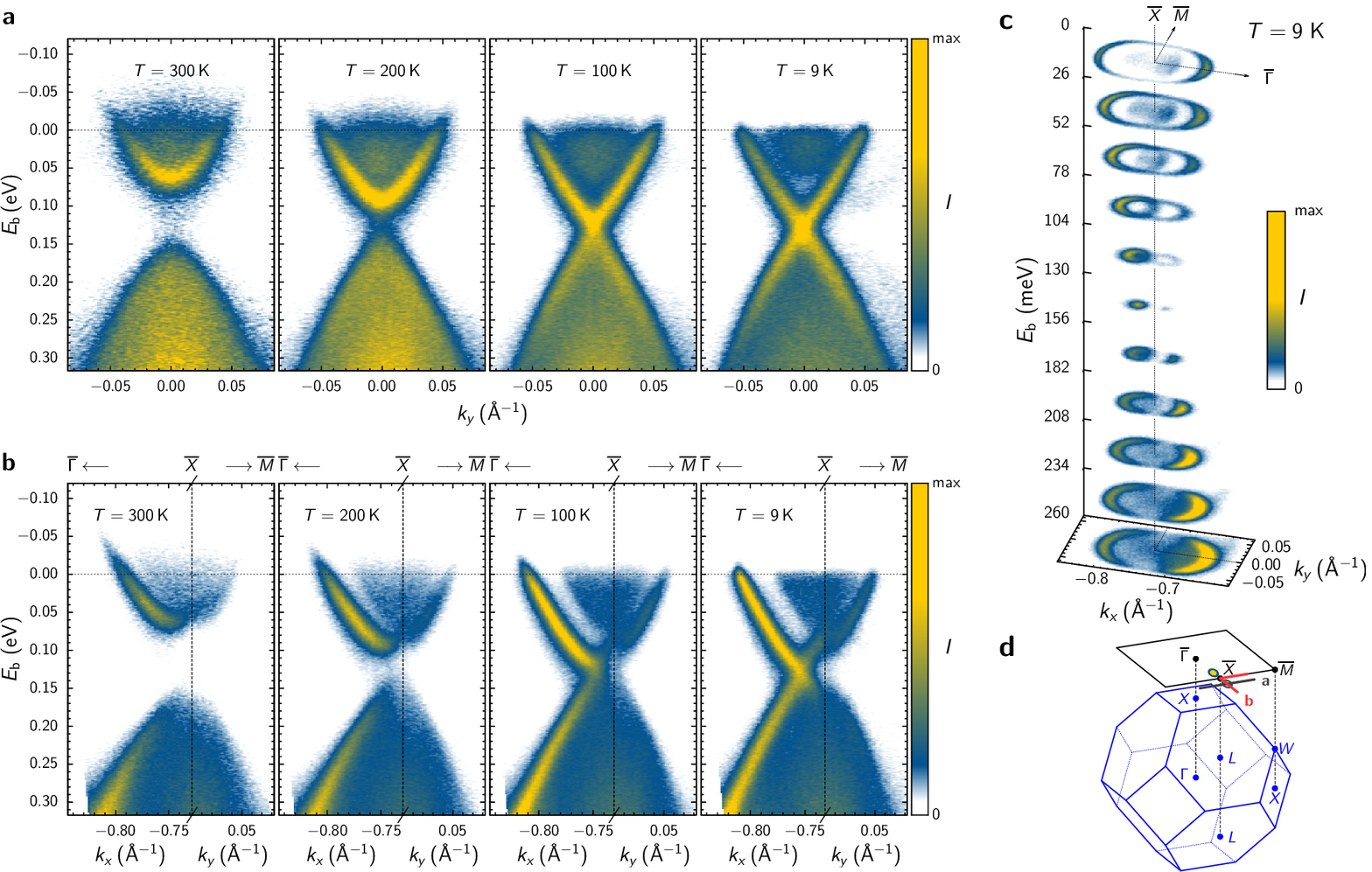}
\caption{{\bf ARPES studies of the (001) surface of \PbSnSeTitle monocrystals.}~\cite{*[{While plotting the data for the published version of this article in }] [{ a sign error occurred. Therefore, the data have been mapped to a wrong $\overline{X}$ point of the (001) surface Brillouin zone. The plots shown here have been corrected. This mistake has no influence on the conclusions of our paper.}] Dziawa-NatMater-2012} \textbf{a}, The temperature dependence of the ARPES spectra in the vicinity of  measured with $10.5$~eV photons. The data are recorded along line $a$ in \textbf{d}. They clearly show the evolution of the gapped surface states (for $T\geq 100$~K) into the Dirac-like state upon lowering the temperature ($T=9$~K). \textbf{b}, The ARPES data along the same $k$-space line as in Figure~\ref{fig:2} ($\overline{\Gamma}$--$\overline{X}$--$\overline{M}$) as indicated in \textbf{d}. \textbf{c}, The corresponding selected constant-energy surfaces close to  at $T=9$~K. \textbf{d}, A sketch of the 3D Brillouin zone and its projection on the (001) surface zone of \PbSnSe. The lines \textbf{a} and \textbf{b} correspond to the data shown in \textbf{a} and \textbf{b}, respectively.}
\label{fig:3}
\end{figure*}

The results of the tight-binding calculations for \PbSnXTe are presented in Figure~\ref{fig:2} for a Sn content of $x=0.2$, $0.381$, $0.6$, and $1$. The $k=0$ value corresponds to the $\overline{X}$ point of the surface Brillouin zone (the projection of the $L(111)$ points in the bulk crystal onto the (001) surface, see Fig.~\ref{fig:3}d). For the critical composition $x_\mathrm{c}=0.381$ (Fig.~\ref{fig:2}b), the crystal bandgap in \PbSnXTe is still slightly open due to the finite thickness of the slab (the confinement effect) as verified by a thickness-dependence analysis shown in the Supplementary Information, section~\ref{sec:S5}. Surface states with dispersion curves notably separated from the dominant `bulk' lines have been identified. For SnTe ($x=1$, Fig.~\ref{fig:2}d), our calculations confirm the key theoretical finding reported by Hsieh et al., i.e. the presence of surface states crossing the bandgap along the $\overline{X}$--$\overline{\Gamma}$ direction~\cite{Hsieh-NatCommun-2012}. The qualitative difference observed between the \PbSnXTe alloy with $x=0.2$ (Fig.~\ref{fig:2}a) and $x=0.6$ (Fig.~\ref{fig:2}c) is the same difference that is expected in a temperature-range-controlled band-inversion experiment using \PbSnSe, i.e., a transition from a $100$~meV bandgap to a $100$~meV inverted bandgap. We note that the dispersive surface states exist for crystals with Sn content below, as well as above the critical composition $x_\mathrm{c}$.

The surface states are observed at the edges of both the valence and the conduction bands. However, only for crystals with an inverted band ordering ($x > x_\mathrm{c}$) we observe the TCI electronic states closing the bandgap in the $\overline{X}$--$\overline{\Gamma}$ direction. Therefore, our calculations confirm that the TCI states are present in the \PbSnXTe and \PbSnXSe alloys when the tin content $x > x_\mathrm{c}$, i.e., when there is an inverted band ordering like the one found in SnTe~\cite{Hsieh-NatCommun-2012}. We note, that this conclusion was reached in the virtual crystal approximation and is expected to be valid for physical effects involving extended Bloch-like 2D and 3D electronic states. A detailed analysis of the influence of local nanoscale electronic disorder, inevitably present in substitutional alloys, on the TCI states (e.g. possible opening of the gap at the Dirac point) is a new intriguing theoretical challenge.

ARPES is a powerful technique for directly probing the electronic structure of solids. Due to its surface sensitivity, this technique is particularly well-suited for studying surface states, and it can therefore be used to verify the existence of TCI surface states in \PbSnXSe. Figure~\ref{fig:3} displays results from the ARPES measurements performed using photons with an energy of $10.5$~eV and at various temperatures on the (001) surface of a \PbSnSe monocrystal (for crystal growth details see Methods and Supplementary Information, section~\ref{sec:S1}). In Figure~\ref{fig:3}a, energy-momentum slices parallel to the $\overline{X}$--$\overline{M}$ direction (line \textbf{a} in Fig.~\ref{fig:3}d) are shown as a function of temperature, indicating the presence of surface states over the complete temperature range. The surface-state's nature is confirmed by photon-energy dependent measurements (for details of ARPES measurements see Methods and Supplementary Information, section~\ref{sec:S2}). Figure~\ref{fig:3}a also shows that upon lowering the temperature, the surface states evolve into an X-shaped state that is similar to the Dirac fermion states that are characteristic for TIs. In the dispersion along the $\overline{\Gamma}$--$\overline{X}$ direction, which is shown to the left of the vertical dashed line in Figure~\ref{fig:3}b, a similar temperature dependence is observed, indicating the formation of a Dirac cone at temperatures below $100$~K, where the Dirac point is located on the $\overline{\Gamma}$--$\overline{X}$ line at $0.023$~\AA{}$^{-1}$ on the near side of $\overline{X}$. Along the $\overline{X}$--$\overline{M}$ direction, the surface state remains gapped at all temperatures, as may be observed in Figure~\ref{fig:3}b to the right of the vertical dashed line. The low-temperature data presented in Figure~\ref{fig:3}b (cut along line \textbf{b} in Fig.~\ref{fig:3}d) are qualitatively well reproduced by the calculated band structure presented in Figure~\ref{fig:2}c, which shows a gapless state along the $\overline{\Gamma}$--$\overline{X}$ direction and a gapped state along the $\overline{X}$--$\overline{M}$ direction. Figure~\ref{fig:3}c presents stacked constant-energy slices for selected equidistant binding energies at $T=9$~K. At this temperature, the $\overline{X}$ point in the surface Brillouin zone is located at ($k_x,\,k_y$) = ($-0.732,\,0$)~\AA{}$^{-1}$, and consequently, the right and left Dirac points, which are observed at $E_\mathrm{b}=130$~meV, belong to the first and second surface Brillouin zones, respectively, as indicated in Figure~\ref{fig:3}d.

Apart from the Dirac-like states discussed above, the spectra shown in Figure~\ref{fig:3} also contain intensity ``inside'' the surface bands. This spectral weight can be related to the bulk conduction and valence bands, both occupied due to the $n$-type nature of the studied sample. These features are discussed in more detail in the Supplementary Information (section~\ref{sec:S2}).

\PbSnXSe alloys are known for high electron mobility (up to about $30000$~cm$^2$/Vs for \PbSnSe at liquid-helium temperatures~\cite{Dixon-PhysRevB-1971}) related to very small effective masses and large dielectric constants that effectively screen the charge scattering centres~\cite{NimtzAndSchlicht, Khokhlov, Dixon-PhysRevB-1971, Melngailis-PhysRevB-1972}. Therefore, this TCI materials system provides a unique platform for experimental studies of magnetotransport phenomena in the high mobility TCI regime.

We experimentally studied the temperature and magnetic field dependence of the longitudinal and transverse components of the magneto-resistivity tensor in both the trivial insulator and the TCI regimes. We found that the Hall electron concentration was only weakly temperature-dependent (Supplementary Information, section~\ref{sec:S3}, Supplementary Fig.~\ref{fig:S5}), as expected for a strongly degenerate semiconductor~\cite{Dixon-PhysRevB-1971, Melngailis-PhysRevB-1972}. The magnetic field characteristics of both the magnetoresistance and the Hall effect display clear nonlinearities, as shown in the Supplementary Information (section~\ref{sec:S4}, Supplementary Fig.~\ref{fig:S6}). Similar anomalies have been observed in bismuth-based TIs~\cite{Qu-Science-2010, Taskin-PhysRevLett-2011, Analytis-NatPhys-2010, Ren-PhysRevB-2010}.

The observed nonlinearity of the Hall resistivity suggests two additive contributions to the conductivity tensor, of which one can be assigned to the bulk conduction (see the Shubnikov-de Haas magnetoresistance osciallations presented in Supplementary Fig.~\ref{fig:S7}). Even though it is tempting to consider the TCI surface state as the second conduction channel, the strong domination of the bulk conductivity in our $n$-type \PbSnSe samples does not permit such an unambiguous identification.

Altogether, our low-temperature photoemission findings provide conclusive evidence that the \PbSnSe crystal is a TCI. By exploiting the temperature dependence of the band inversion, we observe a topological phase transition from a trivial insulator with gapped surface states to a TCI with Dirac-like metallic surface states. This transition occurs at the band-inversion temperature $T_\mathrm{c} \lesssim 100$~K. The doping and tuning possibilities available in \PbSnXSe, as well as the possibility of selectively removing the crystal symmetry protection of the Dirac states by means of crystal distortion, opens the door to considerable future developments. 

\section*{Methods}
The preparation of the highest quality materials was essential for the studies presented here. \PbSnSe single crystals exhibiting a rock-salt structure were grown by the self-selecting vapour growth (SSVG) method under near-equilibrium thermodynamic conditions~\cite{Szczerbakow-ProgCrystalGrowth-2005, Szczerbakow-JCrystalGrowth-1994}, where the specific profile of the temperature field ensures contact between the growing crystal and the crystal's own source material only. This method offers unique advantages in the growth of compositionally uniform solid solutions of high crystal quality and natural (001) facets (Supplementary Information Fig.~\ref{fig:S1}). For our photoemission and magnetotransport studies, parallelepiped samples were cleaved from the bulk crystals along the (001) crystal planes (see the AFM microscopy pictures in Supplementary Information Fig.~\ref{fig:S2}). All samples were prepared such that they possessed $n$-type conductivity, which was performed to achieve both occupied conduction-band and TCI states.

Angle-resolved photoemission spectroscopy studies were carried out on the (001) surface of cleaved single crystals. These studies were performed using linearly polarised light having photon energies between $10.5$~eV and $40$~eV. Temperature-dependent photoemission experiments were performed at the BALTAZAR laser-ARPES facility~\cite{Berntsen-RevSciInstrum-2011} using $10.5$~eV photons. The total energy and crystal-momentum resolution for these measurements were $5$~meV and $0.008$~\AA$^{-1}$, respectively. The single crystal used in the temperature study was cleaved at room temperature under ultra-high-vacuum (UHV) conditions. The photoemission spectra were acquired at selected temperatures in the range of $9$~K to $300$~K. Complementary measurements at higher photon energies and temperatures as low as $110$~K have been conducted using the I4 beam line~\cite{Jensen-NIMA-1997} at the MAX III synchrotron at MAX-lab, Lund University.

\section*{Acknowledgments}
We would like to acknowledge V.~Domukhovski and A.~Reszka for structural and chemical composition analyses of the crystals, J.~Adell for his help during our beamtime at MAX-lab, S.~Safai for her help in numerical calculations, and P.~Kacman for critical reading of the manuscript. In Poland, this work was supported by the European Commission Network SemiSpinNet (PITN-GA-2008-215368) and by the European Regional Development Fund through the Innovative Economy grant (POIG.01.01.02-00-108/09). In Sweden, this work was made possible through support from the Knut and Alice Wallenberg Foundation and the Swedish Research Council.

%

\newpage

\renewcommand{\thesection}{S\arabic{section}}


\section{Crystal growth, structural and chemical characterization}
\label{sec:S1}

\begin{figure}[hb]
\centering
\includegraphics[width=.9\columnwidth]{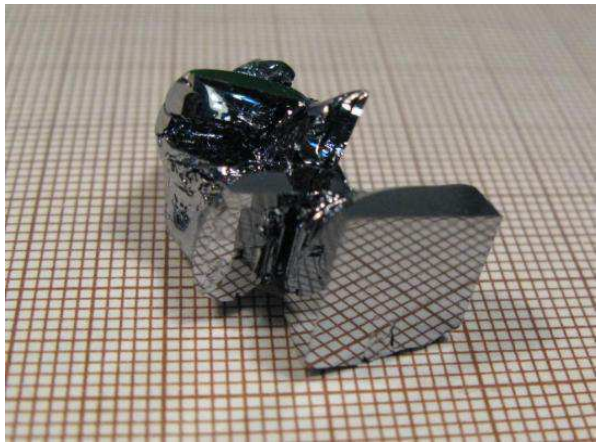}
\caption{{\bf As-grown \PbSnSeTitle bulk monocrystal.} The mirror-like planes at the front part of the crystal are natural (001) facets of a rock-salt-type crystal.}
\label{fig:S1}
\end{figure}

The method of self-selecting vapour growth (SSVG) was applied to manufacture
\PbSnSe monocrystals (Supplementary Fig.~\ref{fig:S1}) of excellent crystal structure as well as
chemical homogeneity. The SSVG method is particularly suitable for optimisation of the
latter because the crystals grow in a near-equilibrium regime. Moreover, the temperature
profile set in the technological furnace ensures growth of a crystal in exclusive contact with
its source material. In the case of the considered alloy, the method allows controlling
deviations from stoichiometry to achieve (without heterodopants) the required $n$-type
conductivity with a moderate electron concentration. The SSVG method was successfully
employed by one of us (A.S.) to grow monocrystals of various II-VI and IV-VI
semiconductor compounds and their solid solutions~\cite{Szczerbakow-ProgCrystalGrowth-2005, Szczerbakow-JCrystalGrowth-1994}.

X-ray diffractometry (XRD) analysis of the crystal structure of \PbSnXSe was done at room
temperature for both as-grown and freshly cleaved (001) crystal surfaces as well as for
powdered polycrystals. It revealed the rock-salt single-crystal phase with a lattice parameter $a=(6.0979 \pm 0.0006)$~\AA{}
corresponding to a Sn content of $x = 0.228 \pm 0.005$~\cite{Szczerbakow-JCrystalGrowth-1994}. XRD rocking-curve measurements of the (002) diffraction peaks yield a full width at half maximum
parameter $\Delta\omega = (70~\mathrm{to}~150)$~arcsec for various (001) surfaces. The chemical composition of
the crystals was determined by energy-dispersive X-ray spectroscopy (EDX) using a scanning
electron microscope SEM Hitachi SU-70 equipped with a Thermo Scientific NSS7
microanalysis system. Probing different places (over an area of $1~\mathrm{mm}^2$) of a sample freshly
cleaved (in air) we found a Sn content of $x = 0.234 \pm 0.005$.

\begin{figure}[hb]
\centering
\includegraphics[width=.9\columnwidth]{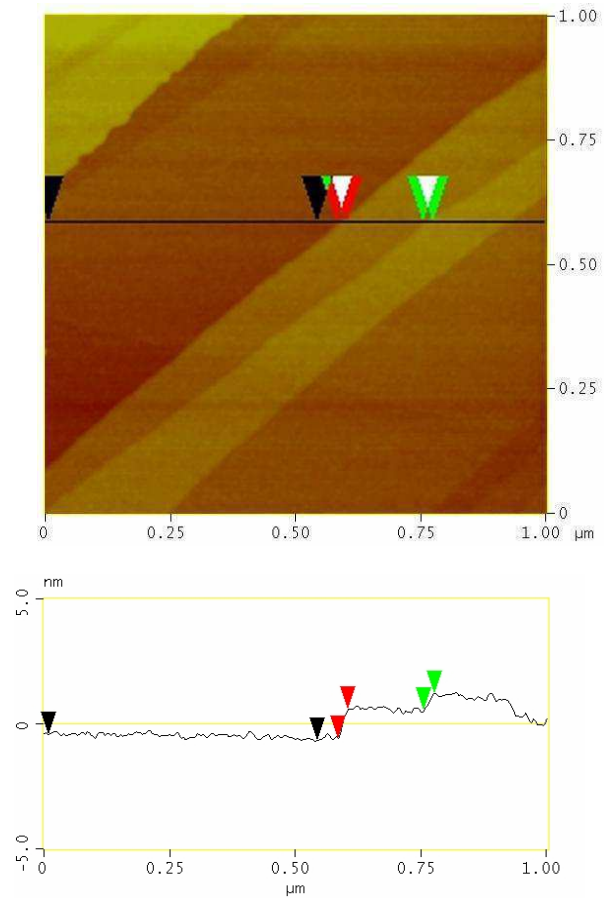}
\caption{{\bf Morphology of the (001) surface of a \PbSnSeTitle crystal.}
Atomic force microscopy (AFM) image of a (001)
surface of a bulk crystal freshly cleaved in air at room
temperature. The height differences between the points
marked with green and red arrows equal $6.8$~\AA{} and $11.5$~\AA{},
respectively. The macroscopically large region between
the black arrows is atomically flat.}
\label{fig:S2}
\end{figure}

The surface morphology of \PbSnSe crystals was studied using a Dimension Icon
Atomic Force Microscope (AFM). The measurements were carried out for (001) surfaces
prepared by cleaving the crystals in air at room temperature. Supplementary Fig.~\ref{fig:S2} presents an
AFM image for an area of $1\times 1~\mu\mathrm{m}^2$ and the section analysis taken along the indicated line.
The root-mean-square surface roughness parameter for this area equals $6.8$~\AA{} (i.e., about just
one lattice parameter). The freshly cleaved (001) surface consists of large atomically flat
regions with small vertical steps with heights of a few monolayers. Time-dependent AFM
measurements of a cleaved surface exposed to air showed substantial surface contaminations
appearing after $1.5$~hours.

\section{Angle-resolved photoemission measurements at various photon energies}
\label{sec:S2}

\begin{figure*}[ht]
\centering
\includegraphics[width=.9\textwidth]{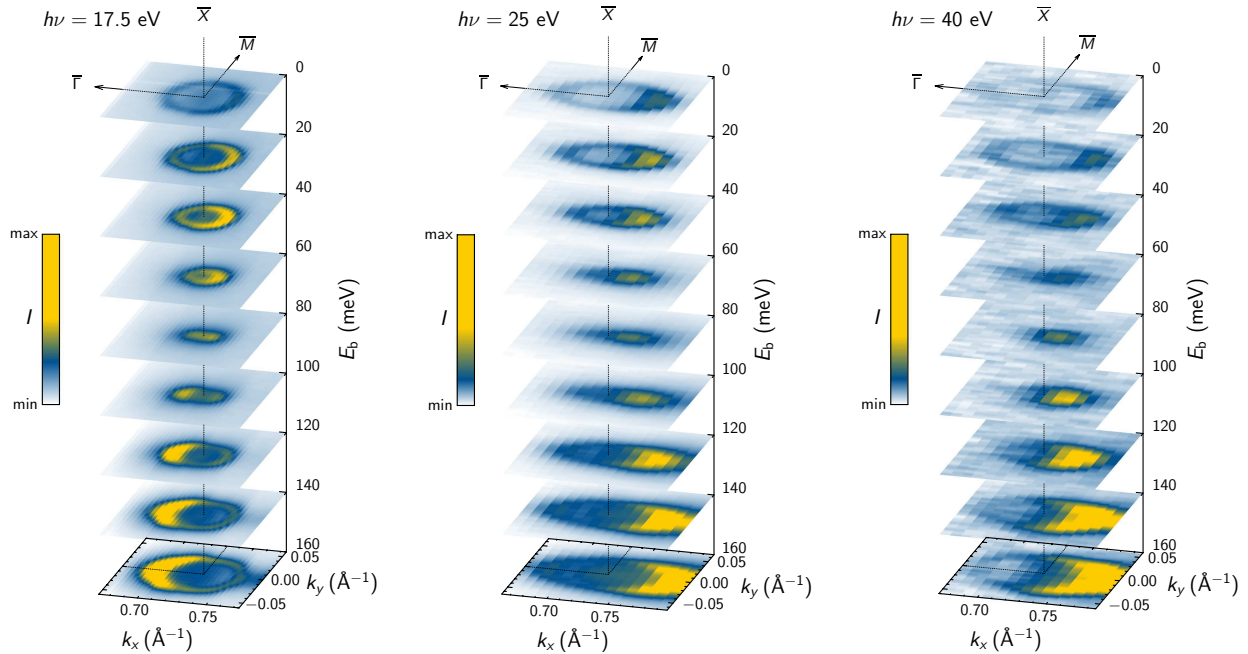}
\caption{Selected constant-energy surfaces of the \PbSnSe surface state for the photon energies $17.5$~eV, $25$~eV and $40$~eV.}
\label{fig:S3}
\end{figure*}

Supplementary Fig.~\ref{fig:S3} shows three sets of selected constant-energy surfaces for \PbSnSe,
in the vicinity of the $\overline{X}$ point, measured at a temperature of $110$~K using photon energies of
$17.5$~eV, $25$~eV, and $40$~eV with corresponding overall energy resolution of $25$~meV to
$40$~meV. The data were acquired at the I4 beam line of the MAX III synchrotron at MAX-lab,
Lund University, Sweden. As seen in the figure, the spectral weight is shifted when changing
the photon energy. For the measurements at $25$~eV and $40$~eV the part of the surface state
located in the second Brillouin zone (BZ) is the most intense. At a photon energy of $17.5$~eV
the most intense part shifts from the first BZ for high binding energies (below the Dirac
points) to the second zone for lower binding energies (above the Dirac points). Apart from
these intensity variations the position of the Dirac-like feature with respect to the surface
Brillouin zone is independent of photon energy (within the limits of experimental accuracy)
suggesting that the observed feature is indeed a surface electronic state.

\begin{figure*}
\centering
\includegraphics[width=.9\textwidth]{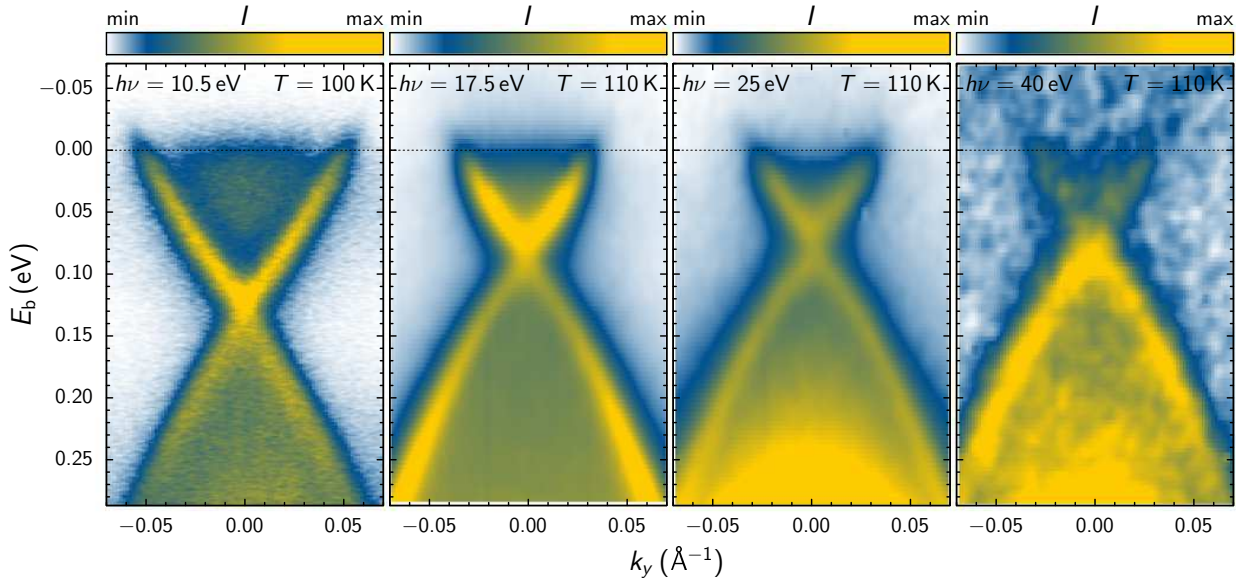}
\caption{Energy-momentum slices across the surface Dirac cones in \PbSnSe taken at the photon energies $10.5$~eV, $17.5$~eV, $25$~eV and $40$~eV.}
\label{fig:S4}
\end{figure*}

Comparing energy-momentum slices from the data sets taken at different photon energies also
confirms that the feature has virtually no photon-energy dependence. Supplementary Fig.~\ref{fig:S4}
displays such energy-momentum slices through the Dirac point located in the second surface BZ
(analogous to Fig.~\ref{fig:3}a in the main text) for the photon energies $10.5$~eV, $17.5$~eV, $25$~eV and
$40$~eV. The cuts are parallel to the $\overline{M}$--$\overline{X}$--$\overline{M}$ direction. The $10.5$~eV data are taken at the
BALTAZAR laser ARPES facility at a temperature of $100$~K and the $17.5$~eV, $25$~eV and $40$~eV data are acquired at the I4 beam line of MAX III at a temperature of $110$~K. Disregarding
for the moment the intensity differences, a clear X-shaped state is seen for all photon
energies. However, for the $17.5$~eV data the Dirac point is shifted about $55$~meV towards
lower binding energy as compared to the $10.5$~eV measurement. A slight further shift towards
lower binding energy is observed for the other measurements. The $10.5$~eV data were
recorded within $12$~hours after cleaving whereas the $17.5$~eV, $25$~eV and $40$~eV data were
recorded at $80$~hours, $85$~hours and $100$~hours post cleaving, respectively. Thus, the shift in
binding energy is likely related to the aging of the surface due to residual gas adsorption.
Similar observations have previously been made for topological insulators, e.g. Bi$_2$Se$_3$, for
which the Dirac point shifts towards higher binding energy upon gas adsorption and these
shifts can amount to several hundred millielectronvolt~\cite{King-PhysRevLett-2011}.
In the present case, the shift is towards lower binding energy and seems
at first glance to be of smaller magnitude. In the Bi$_2$Se$_3$ case, the adsorbtion of residual gas is
accompanied by the appearance of quantum well states that display a large Rashba split~\cite{King-PhysRevLett-2011}. For \PbSnSe, neither the laser-based
data nor the synchrotron data show any clear sign of quantum well states or other
additional surface-related states.

Employing a free-electron model with an inner potential of $15$~eV deduced from normal
emission data, for states close to the Fermi energy at $\overline{X}$ the photon energies $10.5$~eV,
$17.5$~eV, $25$~eV and $40$~eV correspond to $k_z$ values $2.2$, $2.5$, $2.9$ and $3.4$ (in units of $2\pi/a$, where
$a = 6.075$~\AA{}), respectively. However, it should be kept in mind that the $k_z$ resolution in the
studied photon-energy range is very limited as a consequence of the high surface sensitivity.
Nevertheless, different $k_z$ values are probed and bulk-related dispersive features are expected
to differ. The lack of dispersive differences for the Dirac state between the data sets once
more confirms the surface nature of the observed state.

Looking in more detail at the intensity ``inside'' the Dirac state it is clear from Fig.~\ref{fig:3} in the
main text that this intensity in the $10.5$~eV data is partially originating from the ``second
cone'', which has its Dirac point located on the other side of $\overline{X}$. This is not the case for the
higher photon energy measurements since the shift of the chemical potential causes this
contribution to lie above the Fermi level. However, from Supplementary Fig.~\ref{fig:S4} it is apparent
that bulk features contribute to the intensity ``inside'' the cone since it changes as a function of
photon energy. The Fermi level intensity varies with photon energy as does the intensity close
to $k_y = 0$ for high binding energies. The high intensity at the Fermi level in the $17.5$~eV data
and at higher binding energy in the $25$~eV data might indicate that the corresponding points in
reciprocal space are close to a conduction-band minimum and a valence-band maximum,
respectively, although final-state effects and the limited $k_z$ resolution hinder a more accurate
determination.

Returning to the Dirac state, looking at the intensity variations between the different photon
energies one observes that the expected increase in intensity at the Dirac point (c.f. Fig.~\ref{fig:3}a in
the main text), due to the band crossing, is not present. This indicates that the gap is not yet
fully closed and that the transition temperature is slightly below $100$~K. Some of the data seem to
indicate a closed gap but caution is called for since intensity variations as a function of photon
energy might mask a small gap. Such variations are clearly present since at $10.5$~eV the main
intensity is in the part above the Dirac point whereas the opposite is true for $40$~eV. Also, the
intensity at the Dirac point is varying across the photon energies used. The difference in
energy resolution between the laser ARPES and synchrotron measurements provides another
potential explanation for the apparently closed gap in parts of the data.

\section{Electrical properties}
\label{sec:S3}

\begin{figure*}
\centering
\includegraphics[width=.3\textwidth]{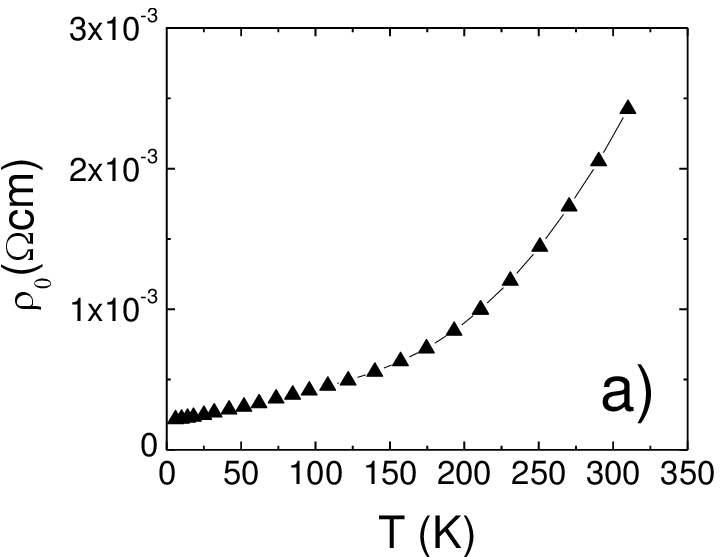}\hfill
\includegraphics[width=.3\textwidth]{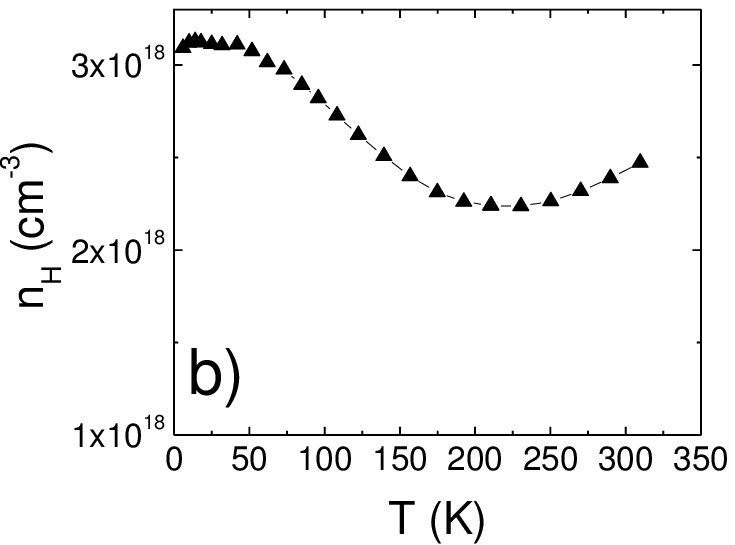}\hfill
\includegraphics[width=.3\textwidth]{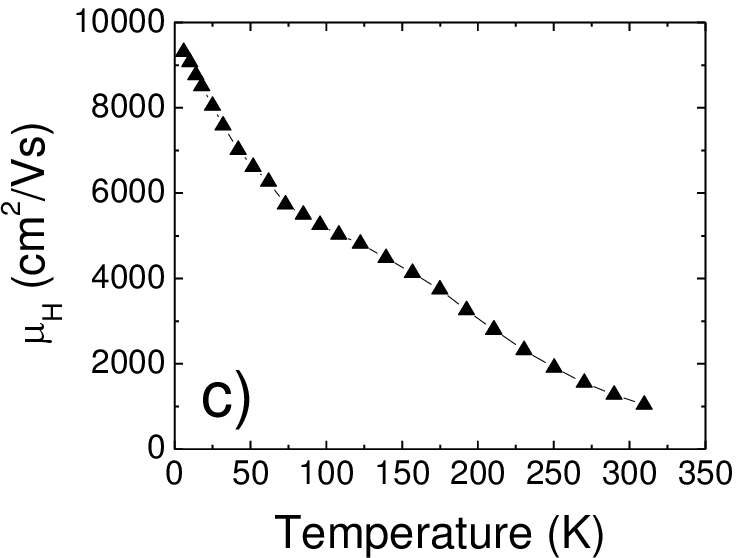}
\caption{{\bf Electrical characterisation of \PbSnSeTitle crystals.}
Temperature dependence of the resistivity (a), the Hall electron concentration (b),
and the Hall electron mobility (c) determined from low-field classical transport measurements.}
\label{fig:S5}
\end{figure*}

The samples for electron transport measurements were prepared from as-grown bulk crystals
by cleaving along three mutually orthogonal $(001)$ planes to get standard parallelepiped Hall-bars
of typical dimensions $2\times 2\times 7~\mathrm{mm}^3$. Directly after cleaving, six contacts were made by
soldering with indium. The low-field ($B = 0.5$~T) Hall effect and resistivity measurements
were performed in a continuous-flow helium cryostat with a DC current of $50$~mA applied
along the $[100]$ crystallographic direction. The electron transport parameters (resistivity, Hall
electron concentration and Hall mobility) presented in Supplementary Fig.~\ref{fig:S5} were obtained
considering one transport channel only in the analysis, as usually done for characterisation
purposes.

\section{High-field magnetotransport measurements}
\label{sec:S4}

\begin{figure*}
\centering
\includegraphics[width=.3\textwidth]{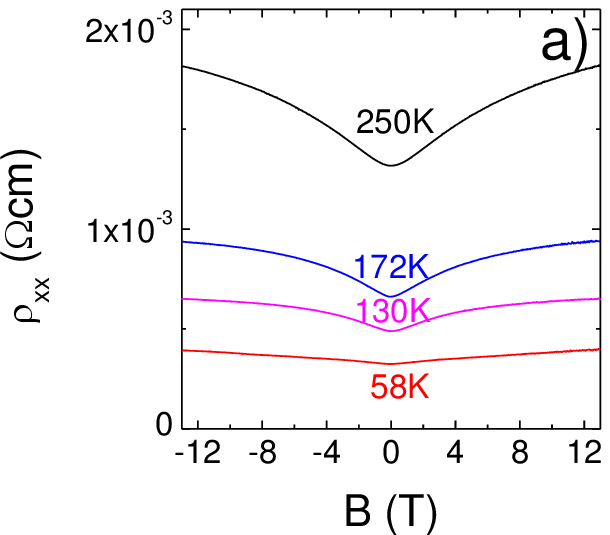}\hfill
\includegraphics[width=.3\textwidth]{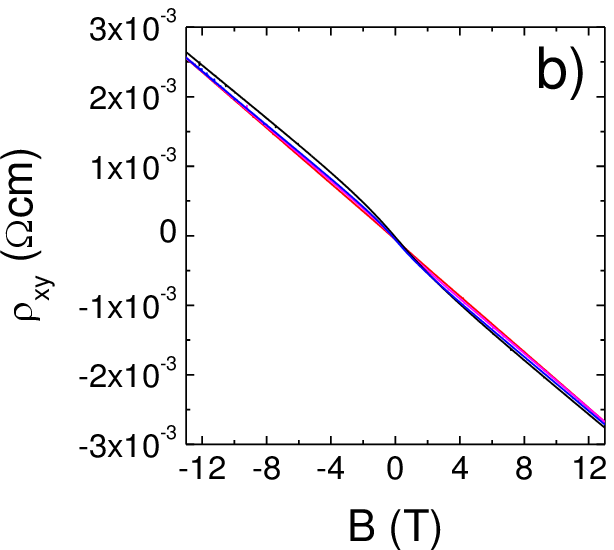}\hfill
\includegraphics[width=.3\textwidth]{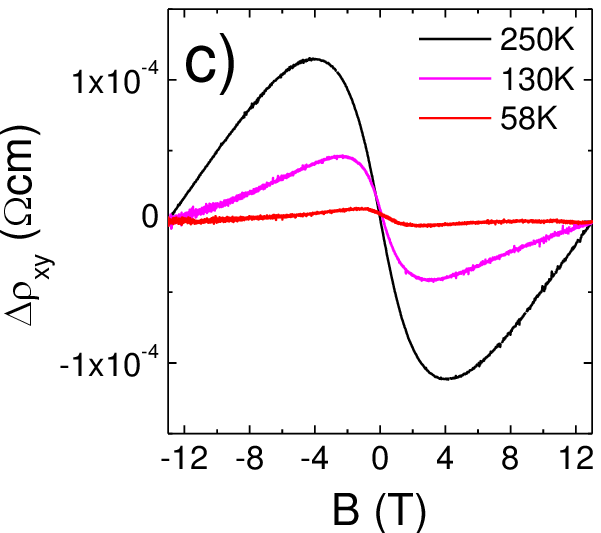}
\caption{{\bf High-field magnetotransport measurements}
The high-field longitudinal resistivity $\rho_{xx}$ (a) and
transverse resistivity $\rho_{xy}$ (b) of the \PbSnSe sample at several temperatures. (c) presents
the $\rho_{xy}$ data after subtraction of the linear term.}
\label{fig:S6}
\end{figure*}

The high-field measurements of the resistivity tensor components $\rho_{xx}$ and $\rho_{xy}$ were carried out
in a liquid-helium cryostat equipped with a $B = 13$~T superconducting solenoid. The
magnetoresistance presented in Supplementary Fig.~\ref{fig:S6}a is of usual orbital origin with
characteristic two-carrier conductivity contributions. Supplementary Fig.~\ref{fig:S6}b displays the as-measured
experimental data for the transverse $\rho_{xy}$ tensor component, which after the
subtraction of linear terms were transformed to the $\Delta\rho_{xy}$ curves shown in Supplementary Fig.~\ref{fig:S6}c.

The experimentally observed magnetic field nonlinearities in the $\rho_{xy}$ tensor component and
the non-quadratic magnetoresistance indicate the presence of at least two parallel conduction
channels. The dominant conduction channel can be identified as a bulk conduction band
contribution. The identification of the other electronic channel needs further magnetotransport
studies, in particular for samples with lower bulk electron concentration.

\begin{figure}
\centering
\includegraphics[width=.9\columnwidth]{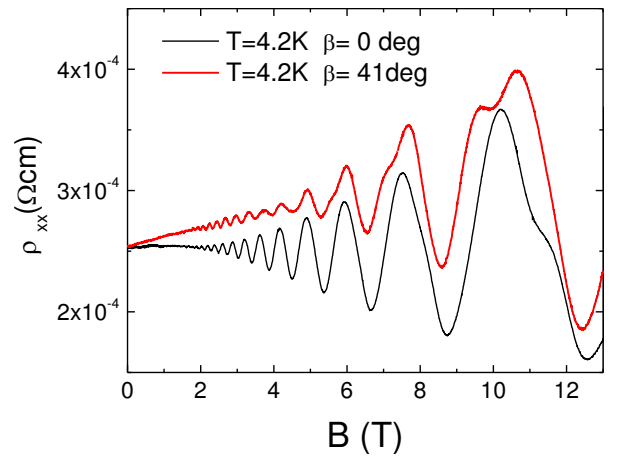}
\caption{{\bf Shubnikov-de Haas oscillations}
The high-field magnetoresistance data recorded for a \PbSnSe sample at the temperature
$T = 4.2$~K in tilted magnetic fields. The tilting angle $\beta$ is the angle between the
external magnetic field and the normal of the $(001)$ sample plane.}
\label{fig:S7}
\end{figure}

The good electrical quality of the \PbSnSe crystals reflected in the high electron mobility
permits the experimental observation at very low temperatures of quantum oscillations of the
magnetoresistance (the Shubnikov-de Haas effect, SdH). The SdH oscillations in $n$-\PbSnSe displayed in
Supplementary Fig.~\ref{fig:S7} are of bulk (3D) origin. This was confirmed
in a tilted-field experiment, where the magnetic-field positions of the oscillations did not
change with the tilting angle. The electron concentration (in a single energy valley) obtained
from the analysis of the period of the SdH oscillations ($n_{\mathrm{SdH}} = 8 \times 10^{17}~\mathrm{cm}^{-3}$)
perfectly corresponds to the total concentration obtained from the low-field Hall effect
($n_{\mathrm{H}} = 3.2 \times 10^{18}~\mathrm{cm}^{-3}$). The factor of $4$ difference is due to the $L$ valley
4-fold degeneracy in \PbSnXSe. Since the bulk contribution dominates the total conductivity at low temperatures,
the SdH oscillations due to topological surface states are not visible in our crystals. The very
small contribution of surface states to the conductivity at low temperatures is typically
observed in other well established topological insulators~\cite{Qu-Science-2010, Taskin-PhysRevLett-2011, Analytis-NatPhys-2010, Ren-PhysRevB-2010}.

\section{Tight-binding (TBA) calculations: the crystal-slab-thickness dependence}
\label{sec:S5}

\begin{figure}[hb]
\centering
\includegraphics[width=.9\columnwidth]{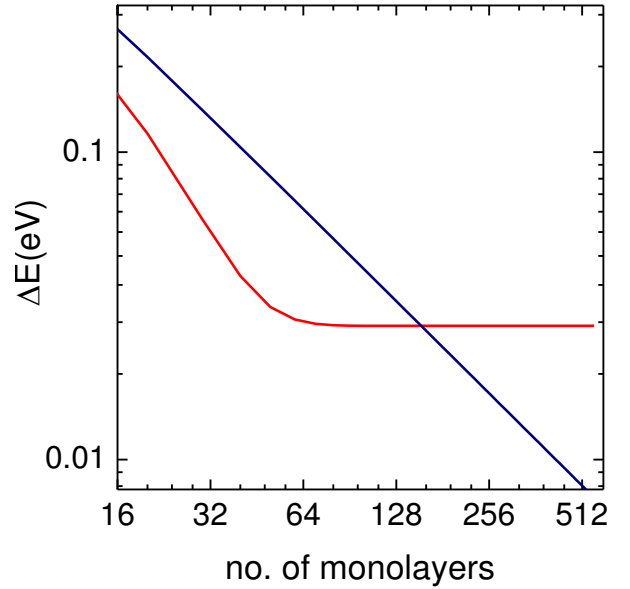}
\caption{{\bf TBA calculations of the crystal-slab-thickness dependence of energy gaps in \PbSnXTeTitle.}
The blue line shows the bandgap at the $\overline{X}$ point in \PbSnXTe with a Sn content $x = x_{\mathrm{c}} = 0.381$, whereas the red line
displays the bandgap on the $\overline{X}$--$\overline{M}$ line for a Sn content $x = 0.6$. For the Sn content range investigated,
the thickness of one monolayer is about $3.2$~\AA{}.}
\label{fig:S8}
\end{figure}

In the case of the critical Sn composition $x_{\mathrm{c}} = 0.381$ the calculated fundamental bulk bandgap
at the $L$ point of the Brillouin zone is closed. However, in our calculations performed for
crystal slabs of finite thickness we observe a small gap at the $\overline{X}$ point. This effect is due to
the electron confinement as verified by the calculation of the bandgap dependence on the
crystal-slab thickness. We have calculated the bandgaps at various points of the surface
Brillouin zone for \PbSnXTe crystal slabs of varying thickness up to about $160$~nm.
Supplementary Fig.~\ref{fig:S8} presents the results in two cases. For the fundamental gap at the $\overline{X}$
point in the crystal with critical Sn content ($x = x_{\mathrm{c}}$) the finite-size effect is clearly observed.
The bandgap is inversely proportional to the slab thickness (blue line), as expected, and
decreases asymptotically to zero with increasing slab thickness. In contrast, the surface states
bandgap in the $\overline{X}$--$\overline{M}$ direction in the crystal with a Sn content $x = 0.6$ (red line) remains
constant for thick slabs thus proving its intrinsic origin. In this case the gap is not determined
by the electron confinement. Its apparent dependence on the thickness for very thin slabs
stems from the surface wave functions overlap.

\end{document}